\documentclass[sigconf]{acmart}

\AtBeginDocument{%
  }

\copyrightyear{2023}
\acmYear{2023}
\setcopyright{rightsretained}
\acmConference[KDD '23]{Proceedings of the 29th ACM SIGKDD Conference on Knowledge Discovery and Data Mining}{August 6--10, 2023}{Long Beach, CA, USA}
\acmBooktitle{Proceedings of the 29th ACM SIGKDD Conference on Knowledge Discovery and Data Mining (KDD '23), August 6--10, 2023, Long Beach, CA, USA}
\acmDOI{10.1145/3580305.3599814}
\acmISBN{979-8-4007-0103-0/23/08}

\usepackage{amsmath,amsfonts}
\usepackage{algorithmic}
\usepackage{graphicx}
\usepackage{xcolor}
\usepackage[linesnumbered,ruled,vlined]{algorithm2e}

\usepackage{booktabs} 
\usepackage{comment}
\usepackage{multirow}
\usepackage{url}
\usepackage{amsmath}
\usepackage{bbold}
\usepackage{balance}

\newcommand{\squishlist}{
 \begin{list}{$\bullet$}
  { \setlength{\itemsep}{0pt}
     \setlength{\parsep}{3pt}
     \setlength{\topsep}{3pt}
     \setlength{\partopsep}{0pt}
     \setlength{\leftmargin}{1.5em}
     \setlength{\labelwidth}{1em}
     \setlength{\labelsep}{0.5em} } }

\newcommand{\squishlisttwo}{
 \begin{list}{$\bullet$}
  { \setlength{\itemsep}{0pt}
     \setlength{\parsep}{0pt}
    \setlength{\topsep}{0pt}
    \setlength{\partopsep}{0pt}
    \setlength{\leftmargin}{2em}
    \setlength{\labelwidth}{1.5em}
    \setlength{\labelsep}{0.5em} } }

\newcommand{\squishend}{
  \end{list}  }

\begin{document}
\title{Empowering Long-tail Item Recommendation through Cross Decoupling Network (CDN)}

\author{Yin Zhang}
\email{yinzh@google.com}
\affiliation{Google Research, Brain Team \country{USA}}

\author{Ruoxi Wang}
\email{ruoxi@google.com}
\affiliation{Google Research, Brain Team \country{USA}}

\author{Tiansheng Yao}
\email{tyao@google.com}
\affiliation{Google Research, Brain Team \country{USA}}

\author{Xinyang Yi}
\email{xinyang@google.com}
\affiliation{Google Research, Brain Team \country{USA}}

\author{Lichan Hong}
\email{lichan@google.com}
\affiliation{Google Research, Brain Team \country{USA}}

\author{James Caverlee}
\email{caverlee@cse.tamu.edu}
\affiliation{Texas A\&M University \country{USA}}

\author{Ed H. Chi}
\email{edchi@google.com}
\affiliation{Google Research, Brain Team \country{USA}}

\author{Derek Zhiyuan Cheng}
\email{zcheng@google.com}
\affiliation{Google Research, Brain Team \country{USA}}

\renewcommand{\shortauthors}{Yin Zhang, et al.}

\begin{abstract}
Industry recommender systems usually suffer from highly-skewed long-tail item distributions where a small fraction of the items receives most of the user feedback. This skew hurts recommender quality especially for the item slices without much user feedback. While there have been many research advances made in academia, deploying these methods in production is very difficult and very few improvements have been made in industry. One challenge is that these methods often hurt overall performance; additionally, they could be complex and expensive to train and serve.

In this work, we aim to improve tail item recommendations while maintaining the overall performance with less training and serving cost. We first find that the predictions of user preferences are biased under long-tail distributions. The bias comes from the differences between training and serving data in two perspectives: 1) the item distributions, and 2) user's preference given an item. Most existing methods mainly attempt to reduce the bias from the item distribution perspective, ignoring the discrepancy from user preference given an item. This leads to a severe forgetting issue and results in sub-optimal performance.

To address the problem, we design a novel Cross Decoupling Network (CDN) to reduce the two differences. Specifically, CDN (i) decouples the learning process of memorization and generalization on the item side through a mixture-of-expert architecture; (ii) decouples the user samples from different distributions through a regularized bilateral branch network. Finally, a new adapter is introduced to aggregate the decoupled vectors, and softly shift the training attention to tail items. Extensive experimental results show that CDN significantly outperforms state-of-the-art approaches on popular benchmark datasets. We also demonstrate its effectiveness by a case study of CDN in a large-scale recommendation system at Google.
\end{abstract}




\maketitle

\section{Introduction}
In industry recommender systems, user feedback towards items usually exhibits server long-tail distributions. That is, a small fraction of items  (\textit{head items}) are extremely popular and receive most of the user feedback, while rest of the items (\textit{tail items}) have very little if any user feedback. Recommender models that are trained based on the long-tail data usually amplify head items, enable the ``rich get richer'' effect while hurting long-term user satisfaction. Models trained on highly skewed data distribution may lead to even worse skewness in real-world applications. Hence, it's critical to address the long-tail distribution problem in industry recommenders.

There have been some methods that are successfully deployed in production models to alleviate the long-tail distribution influence, for example, logQ corrections \cite{yi2019sampling,menon2020long} and re-sampling. However, further improvements in this area for industry models are very limited, despite many research advances made in academia \cite{zheng2021disentangling}. There are several challenges which make putting these research in production models difficult. First, many work targeting long-tail performance would hurt head or overall performance, directly impacting  top line business metrics. Second, production models have very strict latency requirements for real-time inference, while many existing techniques are complex and expensive to serve.  Third, production models prefer simplicity for easy adoption and maintenance. Many research (\emph{e.g.}, meta-learning \cite{lu2020meta}, transfer learning \cite{zhang2021model}) are difficult to be productionized due to these challenges. 

\begin{figure}[t!]
\vspace{-15pt}
\centering  \tiny
\includegraphics[width=2.3in]{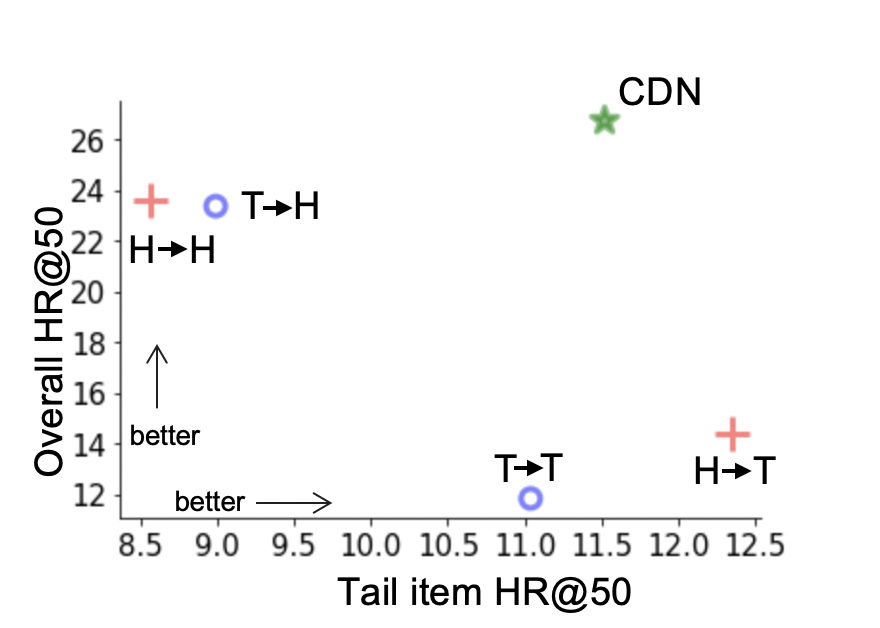}  
\vspace{-15pt}
 \caption{The recommender performance (HR@50) of different methods on tail items (x-axis) and overall items (y-axis). Dots and plus signals represents four two-stage decoupling methods.  `H' means focusing on head items, `T' means focusing on tail items,  and $\rightarrow$ means switching from the 1st to the 2nd stage. When tail (T) is focused in the second stage, the performance on tail items improves; however the overall performance significantly  degrades. Our model CDN achieves excellent performance for both overall and tail item performances. }
 \vspace{-15pt}
 \label{fig:obs}
\end{figure}

In our work, we aim to improve tail item recommendations while maintaining the overall performance with less training and serving cost. We draw our inspiration from recent work with great success in computer vision \cite{kang2019decoupling}. Its core idea is that representation learning and classification learning require different data distributions, where traditionally models don't consider such decoupling for model parameters. Specifically, they propose a two-stage decoupling training strategy, where the first stage trains on the original long-tail distribution for item representation learning, and the second stage trains on the re-balanced data to improve the predictions of tail items. However, in recommendation application, we empirically observe that these methods suffer from a severe \textit{forgetting issue} \cite{toneva2018empirical}. This means that the learned knowledge of certain parts of items (\emph{e.g.} head) in the first training stage are easily forgotten when the learning focus is shifted to other items (\emph{e.g.} tail) in the second training stage, leading to a degradation in overall model quality (as shown in Figure \ref{fig:obs}). Moreover, in large-scale production systems, two-stage training is much more complex to achieve and maintain than co-training scheme. In light of the pros and cons of this method, we target on developing a model that improves the decoupling technique to accommodate web-scale recommenders. 

Like many other methods tackling cold-start problems, the decoupling methods potentially hurt the overall recommendation performance. We attempt to understand this from a theoretical point of view. In particular, we  found that the prediction of user preference towards an item is biased. The bias comes from the differences between training and serving data in two perspectives: 1) the item distributions, and 2) user's preference given an item. Most existing methods mainly attempt to reduce the bias from the item distribution perspective, ignoring the discrepancy from user preference given an item. 


Motivated by the theoretical findings, we propose a novel Cross Decoupling Network (CDN) framework to mitigate the forgetting issue, by considering decoupling from both item and user sides.  ``Decoupling'' means we treat the corresponding learning as two independent processes. In more details:


\squishlist
\item On the item side, the amount of user feedback received by head items and tail items vary significantly. This variations may cause the model to forget the learned knowledge of head (more memorization), when its attention is shifted to tail (more generalization) \footnote{Memorization is to learn and exploit the existing knowledge of visited training data. Generalization, on the other hand, is to explore new knowledge that has not occurred in the training dataset based on the transitivity (\emph{e.g.} data correlation) \cite{cheng2016wide}.}.  Hence, we propose to decouple the memorization and generalization for the item representation learning. In particular, we first group features into memorization and generalization related features, and feed them separately into a memorization-focused expert and a generalization-focused expert, which are then 
aggregated through a frequency-based gating. This mixture-of-expert structure allows us to dynamically balance memorization and generalization abilities for head and tail items.
\item On the user side, we leverage a regularized bilateral branch network to decouple user samples from two distributions. The network consists of two branches: a ``main'' branch that trains on the original distribution for a high-quality representation learning; and a new ``regularizer'' branch that trains on the re-balanced distribution to add more tail information to the model. These two branches share some hidden layers and are jointly trained to mitigate the forgetting issue. Shared tower on the user side are used for scalability. 

\squishend
Finally, a new adapter (called $\gamma$-adapter) is introduced to aggregate the learned vectors from the user and item sides. By adjusting the hyperparameter $\gamma$ in the adapter, we are able to shift the training attention to tail items in a soft and flexible way based on different long-tail distributions.

The resulting model CDN mitigates the forgetting issue of existing models, where it not only improves tail item performance, but also preserves or even improves the overall performance. Further, it adopts a co-training scheme that is easy to maintain. These all make CDN suitable for industrial caliber applications.

The contributions of this paper are 4-fold:
\squishlist
\item We provide a theoretical understanding on how the long-tail distribution influences recommendation performance from both item and user perspectives.
\item We propose a novel cross decoupling network that decouples the learning process of memorization and generalization, and the sampling strategies. A $\gamma$-adapter is utilized to aggregate the learning from the two sides.
\item Extensive experimental results on public dataset show CDN significantly outperforms the SOTA method, improving performance on both overall and tail item recommendations. 
\item We further provide a case study of applying CDN to a large-scale recommender system at Google. We show that CDN is easy to adapt to real setting and achieves significant quality improvements both offline and online.
\squishend


\section{Long-tail Distribution in Recommendation and Motivation} \label{sec:discussion}

\textbf{Problem Settings}. Our goal is to predict user engagement (e.g. clicks, installs) of candidates (e.g. videos, apps) that are in long-tail distributions. We start with formulating the problem: Given a set of users $\mathcal{U} = \{1, 2, \ldots, m\}$, a set of items $\mathcal{I} = \{1, 2, \ldots, n\}$ and their content information (\emph{e.g.} item tags, categories). Let $\hat{d}(u,i)$ be the user $u$'s feedback (e.g. click, install) towards item $i$ in the training set: $\hat{d}(u,i) = 1$ (considered a positive) when $u$ provides a click/install to item $i$, otherwise it is $0$ (considered a negative). Let $d(u,i)$ be the user feedback towards items in the test set, which we aim to predict.


In the training set, the frequency of the user's positive feedback towards item $i$ could be described as $\sum_{u \in \mathcal{U}}\hat{d}(u,i)$, which follows a highly skewed long-tail distribution across items \cite{reed2001pareto}. This means a few head items receive most of the users' feedback, while most items receive little user feedback (tail items). Such imbalance is often quantified by the imbalance factor IF \cite{jamal2020rethinking, lu2020meta} $max_{i, j \in \mathcal{I}} \sum_{u \in \mathcal{U}} \hat{d}(u,i) / \sum_{u \in \mathcal{U}} \hat{d}(u,j)$. 

In industry, IF of item distribution could be very large (\emph{e.g.}, >100000 in production datasets), making it extremely challenging. Furthermore, it also requires the overall performance should be neutral/improvement for production usage. Hence, we target improving recommendation performance for tail items, while maintaining or even improving the overall performance. Key notations are listed in Table \ref{table:notation}.


\begin{table}[t!]
      \caption{Notations.}
      \vspace{-10pt}
       \label{table:notation}
 \begin{center}
    \begin{tabular}{c|l} \toprule
     \textbf{Notation}     &  \textbf{Explanation}   \\
     \hline
     $\mathcal{U}$, $\mathcal{I}$ & user set, item set \\
     $\mathbf{u}$, $\mathbf{i}$ & user $u$ (item $i$) feature vector \\
     $\hat{d}(u,i)$, $d(u,i)$ & user feedback in the traning/test set\\\hline
     $\hat{p}(i|u)$, $p(i|u)$ & given user $u$, the estimated (by training \\&data)/true probability of picking item  $i$ from \\& candidate pool 
     $\mathcal{I}$ \\
    
     $\hat{p}(i)$, $p(i)$ & estimated/true item prior probability \\
     $\hat{p}(i|u)$, $p(u|i)$ & estimated/true conditional probability of\\& user click when item $i$ is known\\\hline
     $\Omega_m$, $\Omega_r$ & training dataset in main/regularizer branch\\
     $\mathbf{x}_m$, $\mathbf{y}_m$ & user/item representation in main branch\\
    $\mathbf{x}_r$, $\mathbf{y}_r$ & user/item representation in regularizer branch \\
    $\alpha$ & learning adapter \\
     \bottomrule
   \end{tabular}
   \end{center}
\vspace{-15pt}
\end{table}

\medskip
\noindent \textbf{Theoretical Analysis and Model Design Motivations}. By Bayes' theorem, we theoretically show the long-tail distribution influence comes from two perspectives:
\squishlist
\item The item distribution discrepancy between training and serving data: the item distribution between the training and serving datasets are often different (a.k.a., training / serving skew), and we refer to this as the discrepancy of the item \textit{prior probability} $p(i)$. This causes the model to have a large bias on serving data, even if the model fits the training data well.
\item The discrepancy of user preference given items between training and serving data: the estimated user preference under the long-tail distribution settings could be biased, and we refer to this as the discrepancy of the \textit{conditional probability} $p(u|i)$. This is especially true for tail items, where the model underfits due to a lack of training examples. This underfitting leads to a large bias even for training data, let alone serving data.  
\squishend

Concretely, for a given user $u$, recommenders aim to accurately approximate the user's preference towards item $i$. The probability of such preference is commonly estimated by the softmax function in recommendation \cite{yi2019sampling}: 
\begin{equation}
p(i | u) = \frac{e^{s(u,i)}}{\sum_{j\in \mathcal{I}}e^{s(u,j)}},
\label{eq:cond}
\end{equation}
where $s(u,i)$ is the true user $u$ preference towards the item $i$, and could be computed through an inner product between user and item embeddings \cite{yi2019sampling}. 

Using Bayes' theorem, we find the estimated $\hat{p}(i|u)$ based on the training set could be described as: 
\begin{align}
\hat{p}(i|u) = \frac{e^{s(u,i) -\log \frac{p(i)}{\hat{p}(i)} -\log \frac{p(u|i)}{\hat{p}(u|i)}}}{\sum_{j \in \mathcal{I}}e^{s(u,j) -\log \frac{p(j)}{\hat{p}(j)} -\log \frac{p(u|j)}{\hat{p}(u|j)}}},
\label{eq:logit}
\end{align}
where $\hat{p}(i)$ and $\hat{p}(u |i)$ are the estimated probabilities learned from the long-tail training dataset. We see that in order for $\hat p(i| u)$ to closely approximate $p(i | u)$ (comparing Eq. \eqref{eq:cond} and Eq. \eqref{eq:logit}), we need both of the estimated prior and conditional probability to well approximate the true probabilities, \emph{i.e.}, $log \frac{p(i)}{\hat{p}(i)} \rightarrow 0$ and $log \frac{p(u | i)}{\hat{p}(u | i)} \rightarrow 0$. Motivated by this, in the following we propose a new model to address the long-tail distribution problem from these two perspectives.

\begin{figure*}[t!]
\centering  \tiny
\vspace{-10pt}
\includegraphics[width=4.6in]{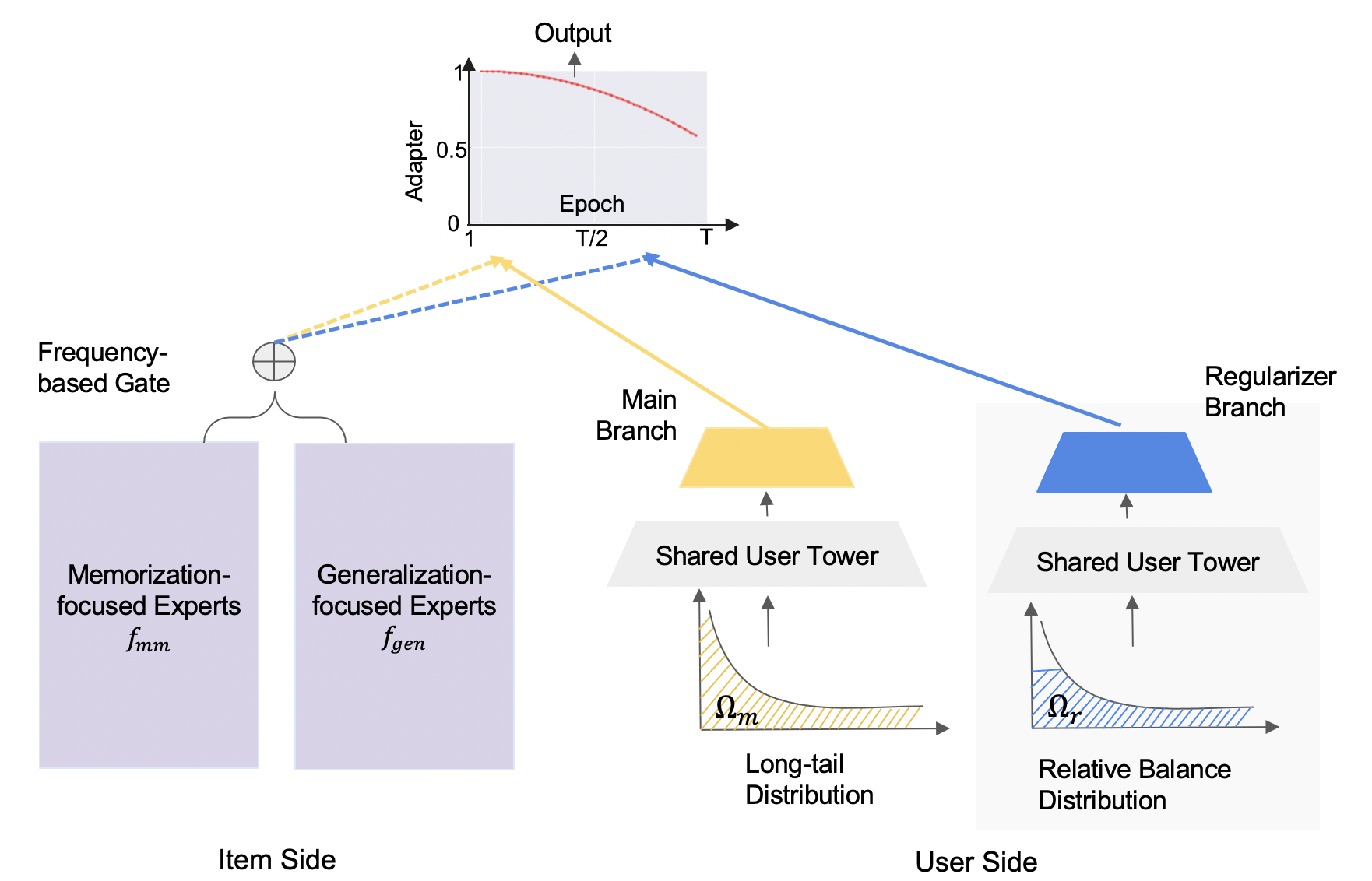}  
\vspace{-10pt}
 \caption{Cross Decoupling Network (CDN).}
 \label{fig:framework} 
 \vspace{-12pt}
\end{figure*}

\section{Cross Decoupling Network}
Based on the analysis, we propose a scalable cross decoupling network (CDN) that addresses the two discrepancies on the item and user sides. The main architecture is depicted in Figure \ref{fig:framework}.
\squishlist
\item On the item side, we propose to decouple the memorization and generalization for head and tail item representation learning. To do so, we utilize a gated mixture of experts (MOE) architecture. In our version of MOE, we feed memorization related features into expert sub networks that focus on memorization. Similarly, we feed content related features into expert sub networks that dedicate for generalization. A gate (\emph{i.e.}, a learnable function) is introduced to decide how much weight the model should put on for memorization and generalization for an item representation. The enhanced item representation learning can mitigate the item distribution discrepancy \cite{jamal2020rethinking}.




\item On the user side, we decouple the user sampling strategies through a regularized bilateral branch network to reduce the user preference discrepancy. The network consists of a main branch for a universal user preference learning, and a regularizer branch to compensate for the sparsity of user feedback towards tail items. A shared tower between two branches are used for scaling up to production usage.


\squishend

Finally, we cross combine the user and item learning to learn the users' diverse preference towards head and tail items in the long-tail distributions with a $\gamma$-adapter.

\subsection{Item Memorization and Generalization Decoupling}


We introduce the concepts of memorization features and generalization features, and then describe our method to decouple them through a gated mixture-of-expert architecture.

\subsubsection{Features for memorization and generalization} \label{sec:features} Industry recommender systems usually consider hundreds of features as model inputs. Instead of encoding those features in the same way, we begin by splitting those features into two groups: memorization features and generalization features. 

\textbf{Memorization features.}
They help memorize the interactions between user and item (collaborative signal) in the training data, such as the item ID. Formally, these features are often categorical features that satisfy:
\squishlist
\item \textbf{Uniqueness}: for its feature space $\mathcal{V}$, $\exists \ f_{in}$ that satisfies $f_{in}$ is an injective function and $f_{in}: \mathcal{I} \rightarrow \mathcal{V}$, and;
\item \textbf{Independence}: for $\forall \ v_1$, $v_2 \in  \mathcal{V}$, the change of $v_1$ does not influence $v_2$;
\squishend
In production recommenders, these features are typically represented by embeddings. Their embedding parameters can only be updated by the corresponding item (uniqueness), and do not share any information with other items (independence). Hence, they only memorize the information for a specific item, and cannot generalize to other existing or unseen items. In the meantime, these features also demonstrate a long-tail distribution due to the uniqueness. Therefore, for features that correspond to head items, their embeddings are updated more often, leading to a remarkable memorization effect. While for features corresponding to tail items, their embeddings may be more noisy due to a lack of gradient update.


\textbf{Generalization features.}
Generalization features are those that can learn the correlations between user preference and item features, and that can be generalized to other items. These features are either shared across different items (\emph{e.g.}, item categories, tags), or are continuous features. Hence, they can generalize to other existing or unseen items, and are important in improving tail item representation learning.


\subsubsection{Item representation learning} We adopt the mixture-of-expert (MoE) structure with a frequency-based gating to decouple the memorization features and generalization features. The structure is depicted on the Item (left) side of Figure \ref{fig:framework}. 

That is, for a training example $(\mathbf{u}, \mathbf{i})$, the item embedding is represented as:
\begin{equation}
    \mathbf{y} = \sum_{k = 1}^{n_1}G(\mathbf{i})_kE_k^{mm}(\mathbf{i}_{mm}) + \sum_{k = n_1+1}^{n_1+n_2}G(\mathbf{i})_kE_k^{gen}(\mathbf{i}_{gen}) \label{eq:itemrep}
\end{equation}
where $E_k^{mm}(\cdot)$ represent the memorization-focused experts which take all memorization features $\mathbf{i}_{mm}$ (\emph{e.g.}, item ID) by concatenating their embeddings as input ; $E_k^{gen}(\cdot)$ represent generalization-focused experts which take all generalization features $\mathbf{i}_{gen}$ (\emph{e.g.}, item categories) by concatenating their embeddings as input; and $G(\cdot)$ is the gating function with $G(\mathbf{i})_k$ being its $k$-th element, and $\sum_{k=1}^{n_1 + n_2}G(\mathbf{i}) = 1 $.
The gating here is critical for dynamically balancing memorization and generalization for head and tail items. Intuitively, the gate could take item frequency as the input, and transform it through a non-linear layer:
$
g(\mathbf{i}) = softmax(\mathbf{W}\mathbf{i}_{freq}),
$
where $\mathbf{W}$ is a learnable weight matrix.
It could also take input from other features, and we found feeding the item popularity as input to work pretty well empirically.

The proposed mechanism uncovers the differences for items in the long-tail distribution in a simple and elegant way for enhanced item representation learning. By decoupling memorization and generalization, the head items can achieve better memorization ability and tail items can be more generalized in the same time. As indicated by \cite{jamal2020rethinking}, the enhanced item representation learning can help compensate the inconsistency of conditional distribution between $p(u|i)$ and $\hat{p}(u|i)$. Additionally, by decoupling the memorization and generalization through experts with frequency-based gates, we are able to mitigate the forgetting issue when the learning attention is shifted towards tail items. That is, with the decoupling, when the training attention shifts to tail items, the gradients (knowledge) from tail items will mainly update the model parameters in the generalization-focused expert, while maintaining the well-learned memorization expert from head items. 


\subsection{User Sample Decoupling}


As shown on the user (right) side of Figure \ref{fig:framework}, inspired by \cite{ren2020balanced,kang2019decoupling,zhang2021model}, we propose a regularized bilateral branch network that consists of two branches: A ``main'' branch that trains on the original highly skewed long-tail distribution $\Omega_m$; and a novel ``regularizer'' branch that trains on a relatively balanced data distribution $\Omega_r$. $\Omega_m$ includes all the user feedback from both head and tail items. $\Omega_r$, on the other hand, includes all the user feedback for the tail items, while down-sampling the user feedback for head items to be as frequent as the most popular tail items. A shared tower between two branches is used for scalability. The method can softly up-weight the learning of user preference towards tail items. So this corrects the under-estimation of user's preference towards tail items, and mitigates the popularity bias of user preference estimation.

In each step, a training example $(\mathbf{u}_m, \mathbf{i}_m) \in \Omega_m$ and $(\mathbf{u}_r, \mathbf{i}_r) \in \Omega_r$, is fed to the main branch and the regularizer branch respectively. Then the user representation vectors are calculated by:
\begin{equation}
\begin{aligned}
\mathbf{x}_m =  h_m(f(\mathbf{u}_m)), \quad
\mathbf{x}_r =  h_r(f(\mathbf{u}_r)).
\label{eq:branchvec}
\end{aligned}
\end{equation}
where $f(\cdot)$ is a sub-network shared for both branches, $h_m(\cdot)$ and $h_r(\cdot)$ are the branch-specific sub-networks. The shared network helps to communicate the learned knowledge from both distributions, and largely reduces the computational complexity. The branch-specific subnetwork learns the unique knowledge for each data distribution (head and tail items).  Therefore, the $\Omega_m$ and $\Omega_r$ are jointly learned to approximate the $\hat{p}(i)$ to $p(i)$, reduce the inconsistency of prior perspective.

The main branch aims to learn high-quality user representations that retain the characteristics of the original distribution, acting as the cornerstone to support the further learning in regularizer branch. As indicated by \cite{kang2019decoupling}, training on original distribution can learn the best and most generalizable representations. The regularizer branch is designed to (1) add more tail information to the model and alleviate the high IF influence on tail items; (2) prevent over-fitting for tail items through a regularized adapter (the adapter details are discussed in Section \ref{sec:crosslearning}).  

In the application to production, the two branches can be trained simultaneously. So the extra training cost would be limited. Note that at inference time, only the main branch is used, so there would be no extra serving cost.

\subsection{Cross Learning} \label{sec:crosslearning}
To bridge the gap between head and tail items, we cross the learning of the decoupled information from user and item side through a $\gamma$-adapter. 

The $\gamma$-adapter is designed to fuse the learned representations and softly shifts the learning focus towards tail items. Specifically, for $\mathbf{x}_m$ and $\mathbf{x}_r$ which are the learned user representations from main and regularizer branch, and $\mathbf{y}_m$ and $\mathbf{y}_r$ which are corresponding learned item representations. The predicted logit is formalized as:
\begin{equation}
s(i_m, i_r) = \alpha_{t} \mathbf{y}^T_m\mathbf{x}_m + (1- \alpha_{t})\mathbf{y}^T_r\mathbf{x}_r,
\end{equation}
where $\alpha_{t}$ is the $\gamma$-adapter, and a function of the training epoch $t$:
\begin{equation}
\alpha_{t} = 1 - (\frac{t}{\gamma \times T})^2, \quad where \ \gamma > 1.
\label{eq:cum}
\end{equation}
Here $T$ is the total number of epochs, and $\gamma$ is the regularizer rate. We see that $\alpha_{t}$ decays as the training progresses (as $t$ increases), and this shifts the model's learning focus from the original distribution to the rebalanced data distribution. In this way, we learn the universal patterns first and then gradually shift towards tail items to improve their performance. This order is critical in obtaining a high-quality representation learning, which can further facilitate the learning of the regularizier branch, as indicated by \cite{zhou2020bbn}. The constrain $\gamma > 1$ is also important in the recommender setting to mitigate the forgetting issue: It ensures that the main focus always stays on the main branch throughout the training. This is a desirable feature for long-tail distributions with different imbalanced factor IF, when IF is high, a larger $\gamma$ is preferred. In fact, we empirically found this $\gamma$-adapter significantly benefits the learning in the highly-skewed long-tail distribution.

With the logit $s(i_m, i_r)$, we calculate the probability of user $u$'s preference towards different items through a softmax function:
\begin{equation}
    p(i|u) = \frac{e^{s(i_m, i_r)}}{\sum_{j \in \mathcal{I}} e^{s(j_m, j_r)}}.
\label{eq:softmax}
\end{equation}
In the industry application, the batch softmax \cite{yi2019sampling} is used for scalability. Then, the loss function can be formulated as:
\begin{equation}
L = -\sum_{u \in \mathcal{U}, i \in \mathcal{I}} \alpha_t \hat{d}(u_m, i_m)\log p(i|u) + (1 - \alpha_t) \hat{d}(u_r, i_r)\log p(i|u),
\label{eq:lg}
\end{equation}
where $\hat{d}(u_m, i_m)$ and $\hat{d}(u_r, i_r)$, respectively, is the user feedback from the main and the regularizer branch. They help to learn high preference scores for items that user engaged with in the two branches. 

For the inference, to predict a user's preference towards an item, we only use the main branch, and calculate the preference score as:
\begin{equation}
    s(u, i) = \mathbf{y}^T_m\mathbf{x}_m.
\label{eq:pred}
\end{equation}
to obtain the logits in softmax. The regularizer branch functions as a regularizer for the training. At prediction time, the test data is also long-tail, and adding the regularizer branch would introduce another layer of mismatch in distributions.

\medskip
\noindent \textbf{Training and serving cost}: CDN has negligible extra cost for training, compared with the standard two tower model. At serving time, the user side only uses the main branch, and the item side has the same total number of parameters/ FLOPs as in the two tower setting. At training time, the extra cost on the user side is on the head for the regularizer branch, which is negligible.

\medskip
\noindent \textbf{Discussion}: An intuitive question is why we decouple the user and item sides from different aspects? In this work, we consider the long-tail distributions from the item side (\emph{i.e.} the long-tail item distribution), which treats users as samples in the long-tail distributions. If we want to consider the long-tail distribution from the user side, then a direct approach would be switching the decoupling methods between users and items as \cite{zhang2021model}. However, we argue that the long-tail user distributions should be modeled differently since the IF for the user side is usually substantially smaller than the item side. Additionally, the long-tail distributions from both sides are highly correlated and they can influence the IF in each side. This is a nontrivial problem and we leave it for future exploration.

\section{Experiment}
In this section, we conduct extensive experiments to address the following key questions:

\textbf{RQ1}: How well does the cross decoupling network CDN perform compared to the state-of-the-art methods, especially  the traditional decoupling methods?

\textbf{RQ2}: How does the cross decoupling work in user and item sides? 

\textbf{RQ3}: How would the expert design influence the CDN performance? How does the model balance between tail/head items by gating mechanisms?

\textbf{RQ4}: What's the influence of $\gamma$-adapter for recommendations on head and tail items?

\textbf{RQ5}: Are we learning better representations for tail items?

\textbf{RQ6}: How does the CDN work on real-world recommenders?

\subsection{Experimental Setup}
\textbf{Datasets.} We use both public benchmark datasets and production dataset at Google which contain rich user and item content information. The public datasets are MovieLens 1M \footnote{https://grouplens.org/datasets/movielens/1m/} and BookCrossing \footnote{http://www2.informatik.uni-freiburg.de/\textasciitilde cziegler/BX/}. The production dataset is from user logs in Google app ads pInstall. Items in three datasets follow the highly-skewed long-tail distribution, which suggests that they are well-suited for studying the targeted long-tail problem. For most experiments, we report the averaged metrics from 5 independent trials followed by the standard error of the mean after the $\pm$ sign.

\medskip
\noindent \textbf{Evaluation Criteria.} We measure the model performance using the widely used metrics \cite{zhang2021model,wang2019neural} -- Hit Ratio at top K (HR@K) and NDCG at top K (NDCG@K). HR@K measures the percentage of whether the ground-truth item is retrieved in the top K ranked items. NDCG@K further considers the ranked positions of the ground-truth items by assigning higher scores if ground-truth items are correctly recommended and have lower ranked positions.

We report both metrics evaluated on the tail, head slices and overall items, to provide different perspectives of recommendation performance. Similar as \cite{zhang2021model}, based on the Pareto Principle \cite{box1986analysis,reed2001pareto},  we consider the top 20\% most frequent items in MovieLens1M and 0.1\% items in BookCrossing \footnote{The split rate of 0.1\% can better show the performance differences among methods, especially for tail items.} as head items, and rest as tail items.


Our goal is to build a single model that can improve the recommendation performance on tail items, while maintaining or even improving the overall performance that can be easily applied for production usage with good scalability and efficiency. 

\medskip
\noindent \textbf{Baselines}. We use the widely adopted production model Two-tower model as the backbone model. Baselines are choosen based on the production usage and similar training/serving time.
\squishlist
\item\textit{Two-tower Model} \cite{yi2019sampling}: The Two-tower model is a popular architecture for modeling different types of content information. Due to its scalability and efficiency, it is widely adopted in real-world applications \cite{zhang2021model,ma2020off,yao2021self}.
\squishend

\noindent Re-balance strategies:
\squishlist
\item\textit{ClassBalance} \cite{cui2019class}:  It calculates an effective number of samples for each imbalanced class and adopts it in a re-weighting term in loss function. We adapt the method to the recommendation task similar to \cite{zhang2021model}. 

\item\textit{LogQ} \cite{yi2019sampling,menon2020long}:  The method is widely applied in production usage. It constructs an item frequency-related term to adjust the softmax logit in the loss function, in order to re-weight head and tail items differently in the learning process.
\squishend

\noindent Decoupling:
\squishlist
\item\textit{NDP} \cite{kang2019decoupling}: The method is adapted from a recent state-of-the-art method for long-tail distribution problem in computer vision. Instead of jointly learning image representation and classification, it separately considers the two learning process in a two-stage training and shows great success in image classification. We adapt the method to recommendation, and treat users as representations and items as classes as \cite{yi2019sampling}. 

\item\textit{BBN} \cite{zhou2020bbn}:  BBN utilizes an uniform sampler and a reversed sampler to create a Bilateral-Branch Network (BBN), and jointly learns the two sampled distributions through an adaptor. In this work, BBN is applied to the image side to improve its representation learning. Hence, we follow a similar strategy where we treat the user as the representation learning and item as the classifier. For cumulative learning, $\alpha_t = 1 - (\frac{t}{T})^2$ is used.

\squishend



\begin{table*}[t!]
\caption{The recommendation performance of CDN versus baselines on MovieLens1M. Numbers after $\pm$ indicate the standard error of the mean. The last row shows the improvement of CDN over the backbone Two-tower model.}
\vspace{-5pt}
\label{table:perfmovie}
\begin{tabular}{c|cc|cc|cc}
\toprule
\multirow{2}{*}{Measure\%}          & \multicolumn{2}{c|}{Overall} & \multicolumn{2}{c|}{Head} & \multicolumn{2}{c}{Tail} \\ \cline{2-7}
          & HR@50        & NDCG@50       & HR@50      & NDCG@50      & HR@50      & NDCG@50     \\\hline
Two-tower &  24.68$\pm$0.16 &  7.29$\pm$0.06 & 33.83$\pm$0.39 & 10.06$\pm$0.14 & 10.32$\pm$0.50 & 2.96$\pm$0.14
  \\ \hline
ClassBalance &20.21$\pm$0.17   &5.91$\pm$0.06  &30.20$\pm$0.20  &8.96$\pm$0.09  &5.19$\pm$0.16  & 1.34$\pm$0.03
\\
LogQ & 13.37$\pm$0.23&	3.50$\pm$0.07&	10.75$\pm$0.38 &	2.62$\pm$0.09&	17.30$\pm$0.15&	4.81$\pm$0.05
\\ \hline
NDP & 14.72$\pm$0.14 & 4.28$\pm$0.06 & 14.92$\pm$0.22 & 4.20$\pm$0.07 & 14.41$\pm$0.22 & 4.41$\pm$0.07
\\
BBN & 16.18$\pm$0.38 &	4.84$\pm$0.17&	22.55$\pm$0.91&	6.89$\pm$0.36 &	6.19$\pm$0.82 &	1.61$\pm$0.25 \\ \hline
\textbf{CDN} &  \textbf{26.75$\pm$0.12} & \textbf{7.93$\pm$0.04} & \textbf{36.46$\pm$0.14} & \textbf{10.97$\pm$0.04} & \textbf{11.51$\pm$0.18} & \textbf{3.15$\pm$0.08} \\ \hline
Improv\%& 8.39\% & 8.78\% & 7.77\% & 9.05\% & 11.53\% & 6.42\%  \\

\bottomrule
\end{tabular}
 \vspace{-5pt}
\end{table*}

\begin{table*}[t!]
\caption{The recommendation performance of CDN versus baselines on BookCrossing. Numbers after $\pm$ indicate the standard error of the mean.  The last row shows the improvement of CDN over Two-tower model.}
\vspace{-5pt}
\label{table:perfbx}
\begin{tabular}{c|cc|cc|cc}
\toprule
\multirow{2}{*}{Measure\%}          & \multicolumn{2}{c|}{Overall} & \multicolumn{2}{c|}{Head} & \multicolumn{2}{c}{Tail} \\ \cline{2-7}
          & HR@50        & NDCG@50       & HR@50      & NDCG@50      & HR@50      & NDCG@50     \\\hline
Two-tower & 2.32$\pm$0.45 &	0.81$\pm$0.17 &	19.76$\pm$4.62 &	7.58$\pm$1.81 &	0.72$\pm$0.08 &	0.19$\pm$0.02
  \\ \hline
ClassBalance &  1.47$\pm$0.32&	0.47$\pm$0.10&	8.94$\pm$2.46&	3.36$\pm$0.88&	0.79$\pm$0.13&	0.21$\pm$0.04
\\
LogQ &1.27$\pm$0.25&	0.37$\pm$0.07&	5.39$\pm$1.54&	1.76$\pm$0.51&	0.89$\pm$0.14&	0.24$\pm$0.04
\\ \hline
NDP &1.58$\pm$0.30&	0.42$\pm$0.08&	7.24$\pm$1.78&	1.94$\pm$0.48&	1.06$\pm$0.19&	0.43$\pm$0.09
\\
BBN &  0.80$\pm$0.20&	0.20$\pm$0.05&	2.62$\pm$0.96&	0.66$\pm$0.26&	0.63$\pm$0.14&	0.16$\pm$0.04  \\ \hline
\textbf{CDN} &  \textbf{2.52$\pm$0.3}  &	\textbf{0.82$\pm$0.15} &	\textbf{21.18$\pm$4.82}  &	\textbf{7.42$\pm$1.86} &	\textbf{0.81$\pm$0.05} &	\textbf{0.22$\pm$0.01}   \\ \hline
Improv\%& 8.94\% & 1.16\% & 7.21\% & -2.09\% & 13.31\% & 12.90\%  \\

\bottomrule
\end{tabular}
 \vspace{-5pt}
\end{table*}

\subsection{Recommendation Performance (RQ1)} 
We first evaluate CDN and baselines on public datasets. All the methods have similar training time. Table \ref{table:perfmovie} and Table \ref{table:perfbx} show the recommendation performance of CDN and baselines on the MovieLens1M and BookCrossing datasets.

The proposed CDN achieves the best performance -- it brings significant improvements not only for tail items, but also for head items and overall recommendation performance. It is worth to notice that most of the baselines only improve on tail items, while the performance of head items significantly degrades (we find inverse propensity score IPS has similar performance as LogQ here). This shows that when we focus the learning attention on tail items, it is very challenging to maintain the performance for head items.


When comparing decoupling methods NDP, BBN and CDN, CDN outperforms the other methods, especially for the overall recommendation. In particular, NDP that adopts the two-stage training improves the recommendation for tail item; however the head item performance degrades. This verifies the forgetting issue of the two-stage training method in recommenders. CDN, on the other hand, is able to mitigate the forgetting issue through the cross decoupling technique, which results in an improvement of not only for tail but also for head items. It is also interesting to note that BBN failed to deliver an improvement either for the tail items or for the head items. One hypothesis is that BBN considers the branch that trains on the balanced data to be as important as the main branch, which does not work well in recommendation where the test distribution is also long tail. This demonstrates the importance to treat the regularizer branch (which takes a balanced distribution) as a regularizer, which softly introduces tail signals to the model while maintaining the main focus on head items. 

Moreover, CDN is more robust for both tail items and overall performance. The standard error of the mean in CDN is smaller than the Two-tower baseline. One possible reason is that CDN is jointly trained with both the original and the balanced dataset. This achieves a similar effect as the data augmentation technique, which effectively alleviates the over-fitting issues for tail items and results in lower model variances.



\subsection{Cross-Decoupling (RQ2)} 
We further did ablation studies to explore how different design choices affect CDN performance:
\squishlist
\item\textit{BDN}: This model considers the main and regularizer branches on the user side to be equally important, as opposed to softly shifting the focus towards regularizer branch and treating it as a regularizer. The preference score is calculated by the averaged outputs of the two branches: $s(i_m, i_r) = 0.5\mathbf{y}^T_m\mathbf{x}_m + 0.5\mathbf{y}^T_r\mathbf{x}_r$.

\item\textit{UDN}: This model only decouples the user side by adopting the Bilateral-Branch Network. For the serving, same as CDN, only the main branch is used to calculate the user preference.

\item\textit{IDN}: This model only decouples the item side by utilizing the memorization-focused and generalization-focused experts with item frequency based gating. The serving is the same as CDN.
\squishend
We see from Table \ref{table:abl} that CDN achieves the best performance for both tail items and head items. Concretely, UDN, IDN and CDN outperforms the two tower model and CDN has the largest improvements. It shows the benefits of cross decoupling. BDN outperforms the two-tower model on the tail items; however the head performance significantly degrades. This verifies the importance of considering the balanced branch as a regularizer rather than a component in the model for the heavily skewed long-tail distribution dataset in recommendation.


\begin{table*}[t!]
\caption{Cross decoupling study on MovieLens1M. Numbers after $\pm$ indicate the standard error of the mean.}
\vspace{-5pt}
\label{table:abl}
\begin{tabular}{c|cc|cc|cc}
\toprule
\multirow{2}{*}{Measure\%}          & \multicolumn{2}{c|}{Overall} & \multicolumn{2}{c|}{Head} & \multicolumn{2}{c}{Tail} \\ \cline{2-7}
          & HR@50        & NDCG@50       & HR@50      & NDCG@50      & HR@50      & NDCG@50     \\\hline
Two-tower &  24.68$\pm$0.16 &  7.29$\pm$0.06 & 33.83$\pm$0.39 & 10.06$\pm$0.14 & 10.32$\pm$0.50 & 2.96$\pm$0.14
  \\ 
BDN & 18.35$\pm$0.21 &	5.27$\pm$0.06&	22.23$\pm$0.37 &	6.31$\pm$0.11 &	12.26$\pm$ 0.17 &	3.64$\pm$ 0.04
\\
UDN & 25.55$\pm$0.08 & 7.53$\pm$0.03 & 34.98$\pm$0.11 & 10.35$\pm$0.06 & 10.77$\pm$0.13 & 3.10$\pm$0.03
\\
IDN & 25.85$\pm$0.15 & 7.58$\pm$0.05 & 34.82$\pm$0.21 & 10.26$\pm$0.08 & 11.79$\pm$0.22 & 3.37$\pm$0.08
\\ 
CDN &  26.75$\pm$0.12 & 7.93$\pm$0.04 & 36.46$\pm$0.14 & 10.97$\pm$0.04 & 11.51$\pm$0.18 & 3.15$\pm$0.08 \\ \bottomrule
\end{tabular}
\vspace{-10pt}
\end{table*}

\subsection{Comparison of Expert Design (RQ3)} 
Is the designed  memorization-focused and generalization-focused experts in CDN can well handle the high heterogeneity between head and tail items in recommendation? To answer the question, we further compared our method with other expert design choices: separate the learning of unbalanced/balanced dataset (similar as the user tower), or head/tail items. We then analyze how head and tail items balance between memorization and generalization through experts' gate values.  

\subsubsection{Recommendation performance} 
To separately consider the head and tail items for item representation learning, there are some design choices: 
\squishlist
\item\textit{Unbalanced/Balanced}: Similar as the user side, we use two experts: one trains on the original dataset, and the other trains on the balanced dataset. Item samples from the two distributions are separately fed to each expert for learning.

\item\textit{Head/Tail}: We directly let one expert trains over head items, and the other trains over tail items. Then, the two experts are aggregated through a 0-1 gating network \cite{ma2018modeling}. 

\item\textit{Mem/Gen}: This is the design in CDN, where one expert considers the memorization features while the other considers the generalization features. 

\squishend

We only change the expert design while keeping the other settings the same for a fair comparison; for different expert designs, model sizes are kept the same. As shown in Table \ref{table:experts}, CDN which adopts Mem/Gen experts achieves the best performance. Specifically, when compared with UDN (in Table \ref{table:abl}) which only decouples the user side, our method that further decouples the item size through Mem/Gen experts shows improvements for both head and tail items; while alternative designs using Unbalanced/Balanced and Head/Tail experts yield a neutral performance for head items. It is worth to note that compared with UDN, the tail performance on Head/Tail relatively degrades. One possible reason is that training tail items alone easily causes over-fitting. This demonstrates the advantages to separately consider the memorization and generalization for item representation learning.




\begin{table*}[t!]
\caption{Expert design study on MovieLens1M. Numbers after $\pm$ indicate the standard error of the mean.}
\vspace{-5pt}
\label{table:experts}
\begin{tabular}{c|cc|cc|cc}
\toprule
\multirow{2}{*}{Measure\%}          & \multicolumn{2}{c|}{Overall} & \multicolumn{2}{c|}{Head} & \multicolumn{2}{c}{Tail} \\ \cline{2-7}
          & HR@50        & NDCG@50       & HR@50      & NDCG@50      & HR@50      & NDCG@50     \\\hline
Unbalanced/Balanced & 25.89$\pm$0.08  &	7.59$\pm$0.02 &	34.69$\pm$0.23 &	10.16$\pm$0.06 &	12.09$\pm$0.32 &	3.57$\pm$0.12
\\
Head/Tail & 25.25$\pm$0.19  &	7.48$\pm$0.06 &	34.64$\pm$0.34 &	10.36$\pm$0.11 &	10.52$\pm$0.41 &	2.97$\pm$0.14
\\ 
Mem/Gen &  26.75$\pm$0.12 & 7.93$\pm$0.04 & 36.46$\pm$0.14 & 10.97$\pm$0.04 & 11.51$\pm$0.18 & 3.15$\pm$0.08 \\ \bottomrule
\end{tabular}
\vspace{-10pt}
\end{table*}

\subsubsection{Memorization vs Generalization} \label{sec:gate}
We then checked how CDN balances memorization and generalization between head and tail items through the learned gate values in CDN. The results are very inspiring and shows many insights.

\begin{figure}[t!]
\centering  \tiny
\includegraphics[width=1.8in]{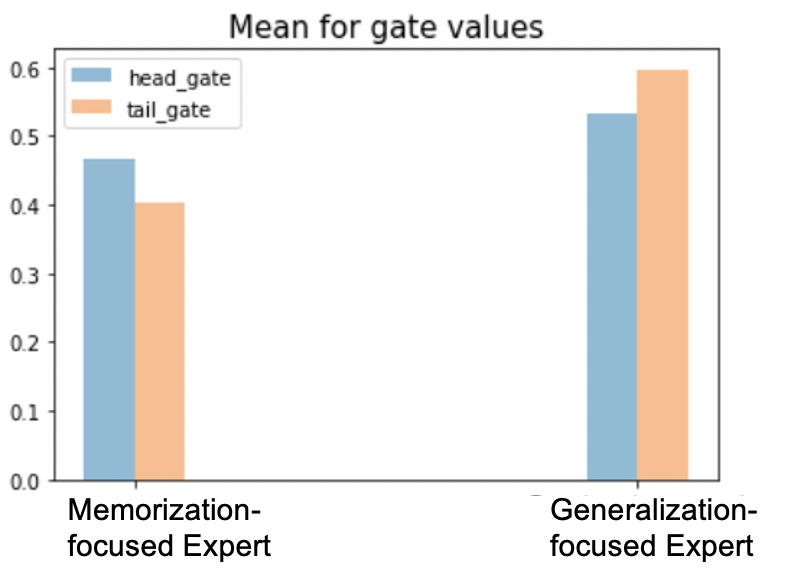}  
\vspace{-10pt}
 \caption{Gate value of memorization-focused and generalization-focused experts for head (blue) and tail (orange) items. In CDN, head items put more weights on memorization-focused experts while tail items put more weights on generalization-focused experts}
 \label{fig:gate} 
\vspace{-10pt}
\end{figure}

Concretely, Figure \ref{fig:gate} shows the gate values of head items (blue bars) and tail items (orange bars) on memorization-focused expert (left) and generalization-focused expert (right). Comparing between values learned from generalization-focused expert (right) and memorization focused expert (left), we see both head and tail items put a higher weight on the generalization-focused expert. It is reasonable since both datasets are very sparse and we utilize rich content information which are better at generalization. More importantly, when comparing between head and tail items, we see that head items put more weights on memorization-focused experts while tail items put more weights on generalization-focused experts. This is consistent with our assumption that head items prefer more memorization abilities as their user feedback information is already very rich, while tail items prefer more generalization abilities due to training data sparsity. This healthy balance of memorization and generalization in CDN is important in improving the performance of both head and tail items.

\subsection{$\gamma$-Adapter (RQ4)} \label{sec:rq4}

The $\gamma$-adapter controls the decay speed of the adaptor function as well as its lower bound. In this section, we check how it influence the recommendation performance on head and tail items.

\begin{figure}[t!]
\centering  \tiny
\includegraphics[width=1.6in]{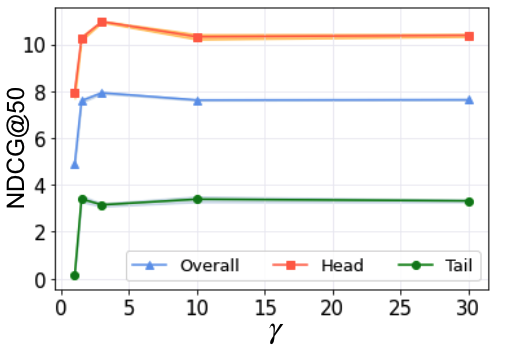}  
 \caption{The NDCG@50 of head, tail and overall performance w.r.t different $\gamma$. }
 \label{fig:gamma} 
\vspace{-10pt}
\end{figure}

Figure \ref{fig:gamma} shows how NDCG@50 changes as we increase $\gamma$ (similar for HR).  When $\gamma$ becomes larger, it puts more weights on the main branch in most epochs. We see in the region where $\gamma$ is small (\emph{e.g.}, $\gamma < 5$ in x-axis), increasing $\gamma$ yields a significant improvement. This suggests the importance of maintaining a focus on the main branch. That is, treating the regularizer branch as a regularization term. When $\gamma$ is large (\emph{e.g.}, $\gamma > 5$), we see that the performance saturated when we further increase $\gamma$. Furthermore, for tail items (green line), larger $\gamma$ could hurt its performance (e.g. when $\gamma$ = 4). This shows the regularizer branch benefits tail item to avoid overfitting. 

\subsection{Representation Visualization (RQ5)} 
\begin{figure}[t!]
\centering  \tiny
\includegraphics[width=2.7in]{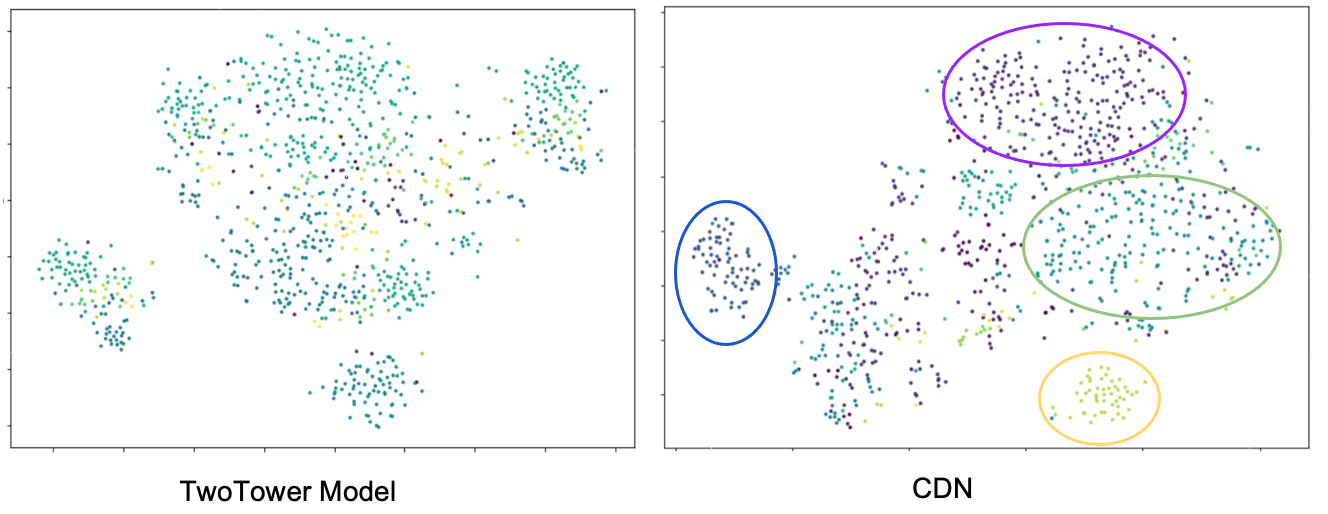}  
\vspace{-10pt}
 \caption{Embedding visualization comparison by using t-SNE \cite{van2014accelerating}. Each genre is represented by a different color.}
 \label{fig:embvis} 
\end{figure}

We visualize the learned embeddings from CDN and two-tower model to analyze semantics of the learnt representations, as shown in Figure \ref{fig:embvis}. We see that compared to the two-tower model, CDN's embeddings are more coherently clustered (movies in the same genre tend to cluster together). This suggests that CDN is able to learn more semantic meaningful representations for tail items, which also aligns with the findings in Section \ref{sec:gate} where tail slice items put more weights on the generalization experts.

\subsection{Experiments in a Google System (RQ6)} 

We conduct a study to showcase CDN framework's performance in a real-world recommender system at Google. Results show that CDN is easy to adapt with industry scale and constraints, and delivers significant offline and online performance through live A/B experiments.


Concretely, given queries and candidates, the task is to predict the likelihood of a user installing an app. The training data is sampled from user logs and consists of hundreds of billions of training examples. Rich content features of query and candidates are used as inputs in the model. We use a binary label of whether a user installed the app, and the positive labels are sparse compared to the negatives. Due to the constraints on the infrastructure side, here we focus on the item side decoupling. And instead of focusing on head and tail items, we adopted different approaches to define dense and sparse data slices based on the production application. Results are shown in table \ref{table:prodlive}.  

\begin{table}[t!]
\caption{Relative improvements of CDN performance.}
\vspace{-5pt}
\label{table:prodlive}
\begin{tabular}{c|ccc}\toprule 
Metrics & Overall      & Dense    & Sparse             \\ \hline
(Offline) AUCLoss improvement & +0.27\% & +0.26\%& +0.66\%\\
(Online) KPI online & +1.43\% & +0.73\%& +3.75\%\\\hline
\end{tabular}
\vspace{-15pt}
\end{table}

As shown in the table \ref{table:prodlive}, CDN delivers significant improvements on top of the production model for both offline and online experiments. We use AUCLoss (1 - AUC) as the offline evaluation metric, and see 0.27\% overall improvements, 0.66\% improvement on the sparse slice, and 0.26\% improvement on the dense slice (0.1\% is considered significant). In a time frame of 14 days, the online metrics were also highly positive, with 1.43\% improvements on key online metric. What is encouraging is that we see even bigger improvements for sparse slice compared to the dense slice, with 3.75\% improvements on key online metric. The results further confirm the effectiveness of CDN to improve the sparse slice performance and model generalization.

\section{Related Work}


\smallskip \noindent
\textbf{Long-tail Problem.} Large-scale real world datasets often exhibit a long-tail distribution, 
that pose critical challenges to many industry recommender systems \cite{niu2020dual,ge2022toward}. Industry solutions usually focus on rebalancing the data distribution through resampling and reweighting \cite{cui2019class,yi2019sampling,menon2020long}.
While improving performance for tail classes, they may under-represent the head classes. In fact, recent studies \cite{kang2019decoupling, zhou2020bbn} find that the rebalancing methods could hurt the representative capability when the original distribution gets distorted badly.

Many of the progresses mentioned above are achieved in academia \cite{dahiya2019deepxml,cao2021learning}, especially computer vision area \cite{kang2019decoupling,wang2020long,zhang2021test}. However, directly adapting those methods to industry-scale recommenders is very challenging. For example, a closely related academia work \cite{zhang2021model} uses meta-learning to address the long-tail distribution problem. It shows great success in the academia datasets. However, we find the training time of the method to be doubled compared to the production baseline. The increased training time cannot justify the performance gain, making it infeasible in practice. Furthermore, in recommendation, the imbalance factor IF is magnitudes higher, and IF is a core factor that influence the recommendation performance. Moreover, to improve the performance of tail items, most related work in recommendation are focusing on cold-start items by investigating effective ways to incorporate content information \cite{zhou2011functional,lu2020meta}. This is a different scenario from the long-tail distribution problem. In our long-tail case, we need to consider the influence from the imbalanced distribution and aim to not only improve the tail performance but also maintain the overall performance. 

\smallskip \noindent
\textbf{Memorization and Generalization.} This is a key topic that has been widely studied \cite{zhang2019identity,brown2021memorization,arpit2017closer}. They directly influence the model's expressiveness and performance on unseen data or new tasks. \cite{feldman2020does} theoretically shows that memorizing data labels is necessary to achieve a good performance, and many experiments \cite{feldman2020neural} further strengthen the importance of memorization. Recently, generalization receives extensive attention for different model designs, such as meta-learning \cite{fallah2021generalization}, reinforcement learning \cite{wang2020improving}. Research has shown that generalization is especially crucial when the data is sparse or insufficient for a given task, \emph{e.g.}, the distribution shift and adversarial attack \cite{zhu2021understanding} problems. \cite{zhang2019identity} conducted empirical analysis on the generalization for different models, and shows that a model's generalization ability is closely related to its architecture. In recommendation, the wide-and-deep model \cite{cheng2016wide} explains the importance of considering both memorization and generalization in model design, and is widely adopted in many real world applications. 




\section{Conclusion}
In this work, we propose a scalable cross decoupling network (CDN) to address the long-tail distribution in industry recommender systems. CDN decouples the memorization and generalization on the item side, as well as the user sampling from the user side. Then, a new $\gamma$-adapter is introduced to softly shift the learning attention to tail items, while smoothly aggregating the learning from both sides. Experiments on public dataset show that CDN delivers significant improvements for both tail and overall recommendations. The case study of applying CDN to a large-scale recommendation system at Google further demonstrates CDN can also improve overall model performance and model generalization in real world scenarios.

\section*{Acknowledgments}
We would like thank our amazing collaborators for their collaboration and highly valuable suggestions, including Ting Chen, Evan Ettinger, Andrew Evdokimov, and Samuel Ieong.

\bibliographystyle{ACM-Reference-Format}
\balance
\bibliography{ref}


\begin{thebibliography}{34}


\ifx \showCODEN    \undefined \def \showCODEN     #1{\unskip}     \fi
\ifx \showDOI      \undefined \def \showDOI       #1{#1}\fi
\ifx \showISBNx    \undefined \def \showISBNx     #1{\unskip}     \fi
\ifx \showISBNxiii \undefined \def \showISBNxiii  #1{\unskip}     \fi
\ifx \showISSN     \undefined \def \showISSN      #1{\unskip}     \fi
\ifx \showLCCN     \undefined \def \showLCCN      #1{\unskip}     \fi
\ifx \shownote     \undefined \def \shownote      #1{#1}          \fi
\ifx \showarticletitle \undefined \def \showarticletitle #1{#1}   \fi
\ifx \showURL      \undefined \def \showURL       {\relax}        \fi
\providecommand\bibfield[2]{#2}
\providecommand\bibinfo[2]{#2}
\providecommand\natexlab[1]{#1}
\providecommand\showeprint[2][]{arXiv:#2}

\bibitem[Arpit et~al\mbox{.}(2017)]%
        {arpit2017closer}
\bibfield{author}{\bibinfo{person}{Devansh Arpit},
  \bibinfo{person}{Stanis{\l}aw Jastrz{\k{e}}bski}, \bibinfo{person}{Nicolas
  Ballas}, \bibinfo{person}{David Krueger}, \bibinfo{person}{Emmanuel Bengio},
  \bibinfo{person}{Maxinder~S Kanwal}, \bibinfo{person}{Tegan Maharaj},
  \bibinfo{person}{Asja Fischer}, \bibinfo{person}{Aaron Courville},
  \bibinfo{person}{Yoshua Bengio}, {et~al\mbox{.}}}
  \bibinfo{year}{2017}\natexlab{}.
\newblock \showarticletitle{A closer look at memorization in deep networks}. In
  \bibinfo{booktitle}{\emph{ICML}}. PMLR, \bibinfo{pages}{233--242}.
\newblock


\bibitem[Box and Meyer(1986)]%
        {box1986analysis}
\bibfield{author}{\bibinfo{person}{George~EP Box} {and}
  \bibinfo{person}{R~Daniel Meyer}.} \bibinfo{year}{1986}\natexlab{}.
\newblock \showarticletitle{An analysis for unreplicated fractional
  factorials}.
\newblock \bibinfo{journal}{\emph{Technometrics}} \bibinfo{volume}{28},
  \bibinfo{number}{1} (\bibinfo{year}{1986}).
\newblock


\bibitem[Brown et~al\mbox{.}(2021)]%
        {brown2021memorization}
\bibfield{author}{\bibinfo{person}{Gavin Brown}, \bibinfo{person}{Mark Bun},
  \bibinfo{person}{Vitaly Feldman}, \bibinfo{person}{Adam Smith}, {and}
  \bibinfo{person}{Kunal Talwar}.} \bibinfo{year}{2021}\natexlab{}.
\newblock \showarticletitle{When is memorization of irrelevant training data
  necessary for high-accuracy learning?}. In
  \bibinfo{booktitle}{\emph{Proceedings of the 53rd annual ACM SIGACT symposium
  on theory of computing}}. \bibinfo{pages}{123--132}.
\newblock


\bibitem[Cao et~al\mbox{.}(2021)]%
        {cao2021learning}
\bibfield{author}{\bibinfo{person}{Yixin Cao}, \bibinfo{person}{Jun Kuang},
  \bibinfo{person}{Ming Gao}, \bibinfo{person}{Aoying Zhou},
  \bibinfo{person}{Yonggang Wen}, {and} \bibinfo{person}{Tat-Seng Chua}.}
  \bibinfo{year}{2021}\natexlab{}.
\newblock \showarticletitle{Learning relation prototype from unlabeled texts
  for long-tail relation extraction}.
\newblock \bibinfo{journal}{\emph{IEEE Transactions on Knowledge and Data
  Engineering}} (\bibinfo{year}{2021}).
\newblock


\bibitem[Cheng et~al\mbox{.}(2016)]%
        {cheng2016wide}
\bibfield{author}{\bibinfo{person}{Heng-Tze Cheng}, \bibinfo{person}{Levent
  Koc}, \bibinfo{person}{Jeremiah Harmsen}, \bibinfo{person}{Tal Shaked},
  \bibinfo{person}{Tushar Chandra}, \bibinfo{person}{Hrishi Aradhye},
  \bibinfo{person}{Glen Anderson}, \bibinfo{person}{Greg Corrado},
  \bibinfo{person}{Wei Chai}, \bibinfo{person}{Mustafa Ispir}, {et~al\mbox{.}}}
  \bibinfo{year}{2016}\natexlab{}.
\newblock \showarticletitle{Wide \& deep learning for recommender systems}. In
  \bibinfo{booktitle}{\emph{Proceedings of the 1st workshop on deep learning
  for recommender systems}}. \bibinfo{pages}{7--10}.
\newblock


\bibitem[Cui et~al\mbox{.}(2019)]%
        {cui2019class}
\bibfield{author}{\bibinfo{person}{Yin Cui}, \bibinfo{person}{Menglin Jia},
  \bibinfo{person}{Tsung-Yi Lin}, \bibinfo{person}{Yang Song}, {and}
  \bibinfo{person}{Serge Belongie}.} \bibinfo{year}{2019}\natexlab{}.
\newblock \showarticletitle{Class-balanced loss based on effective number of
  samples}. In \bibinfo{booktitle}{\emph{Proceedings of the IEEE/CVF conference
  on computer vision and pattern recognition}}. \bibinfo{pages}{9268--9277}.
\newblock


\bibitem[Dahiya et~al\mbox{.}(2019)]%
        {dahiya2019deepxml}
\bibfield{author}{\bibinfo{person}{Kunal Dahiya}, \bibinfo{person}{Anshul
  Mittal}, \bibinfo{person}{Deepak Saini}, \bibinfo{person}{Kushal Dave},
  \bibinfo{person}{Himanshu Jain}, \bibinfo{person}{Sumeet Agarwal}, {and}
  \bibinfo{person}{Manik Varma}.} \bibinfo{year}{2019}\natexlab{}.
\newblock \showarticletitle{DeepXML: Scalable \& Accurate Deep Extreme
  Classification for Matching User Queries to Advertiser Bid Phrases}.
\newblock  (\bibinfo{year}{2019}).
\newblock


\bibitem[Fallah et~al\mbox{.}(2021)]%
        {fallah2021generalization}
\bibfield{author}{\bibinfo{person}{Alireza Fallah}, \bibinfo{person}{Aryan
  Mokhtari}, {and} \bibinfo{person}{Asuman Ozdaglar}.}
  \bibinfo{year}{2021}\natexlab{}.
\newblock \showarticletitle{Generalization of Model-Agnostic Meta-Learning
  Algorithms: Recurring and Unseen Tasks}. In
  \bibinfo{booktitle}{\emph{NeurIPS}}.
\newblock


\bibitem[Feldman(2020)]%
        {feldman2020does}
\bibfield{author}{\bibinfo{person}{Vitaly Feldman}.}
  \bibinfo{year}{2020}\natexlab{}.
\newblock \showarticletitle{Does learning require memorization? a short tale
  about a long tail}. In \bibinfo{booktitle}{\emph{Proceedings of the 52nd
  Annual ACM SIGACT Symposium on Theory of Computing}}.
  \bibinfo{pages}{954--959}.
\newblock


\bibitem[Feldman and Zhang(2020)]%
        {feldman2020neural}
\bibfield{author}{\bibinfo{person}{Vitaly Feldman} {and}
  \bibinfo{person}{Chiyuan Zhang}.} \bibinfo{year}{2020}\natexlab{}.
\newblock \showarticletitle{What neural networks memorize and why: Discovering
  the long tail via influence estimation}.
\newblock \bibinfo{journal}{\emph{arXiv preprint arXiv:2008.03703}}
  (\bibinfo{year}{2020}).
\newblock


\bibitem[Ge et~al\mbox{.}(2022)]%
        {ge2022toward}
\bibfield{author}{\bibinfo{person}{Yingqiang Ge}, \bibinfo{person}{Xiaoting
  Zhao}, \bibinfo{person}{Lucia Yu}, \bibinfo{person}{Saurabh Paul},
  \bibinfo{person}{Diane Hu}, \bibinfo{person}{Chu-Cheng Hsieh}, {and}
  \bibinfo{person}{Yongfeng Zhang}.} \bibinfo{year}{2022}\natexlab{}.
\newblock \showarticletitle{Toward Pareto efficient fairness-utility trade-off
  in recommendation through reinforcement learning}. In
  \bibinfo{booktitle}{\emph{Proceedings of the fifteenth ACM international
  conference on web search and data mining}}. \bibinfo{pages}{316--324}.
\newblock


\bibitem[Jamal et~al\mbox{.}(2020)]%
        {jamal2020rethinking}
\bibfield{author}{\bibinfo{person}{Muhammad~Abdullah Jamal},
  \bibinfo{person}{Matthew Brown}, \bibinfo{person}{Ming-Hsuan Yang},
  \bibinfo{person}{Liqiang Wang}, {and} \bibinfo{person}{Boqing Gong}.}
  \bibinfo{year}{2020}\natexlab{}.
\newblock \showarticletitle{Rethinking class-balanced methods for long-tailed
  visual recognition from a domain adaptation perspective}. In
  \bibinfo{booktitle}{\emph{Proceedings of the IEEE/CVF Conference on Computer
  Vision and Pattern Recognition}}. \bibinfo{pages}{7610--7619}.
\newblock


\bibitem[Kang et~al\mbox{.}(2020)]%
        {kang2019decoupling}
\bibfield{author}{\bibinfo{person}{Bingyi Kang}, \bibinfo{person}{Saining Xie},
  \bibinfo{person}{Marcus Rohrbach}, \bibinfo{person}{Zhicheng Yan},
  \bibinfo{person}{Albert Gordo}, \bibinfo{person}{Jiashi Feng}, {and}
  \bibinfo{person}{Yannis Kalantidis}.} \bibinfo{year}{2020}\natexlab{}.
\newblock \showarticletitle{Decoupling representation and classifier for
  long-tailed recognition}.
\newblock \bibinfo{journal}{\emph{ICLR}} (\bibinfo{year}{2020}).
\newblock


\bibitem[Lu et~al\mbox{.}(2020)]%
        {lu2020meta}
\bibfield{author}{\bibinfo{person}{Yuanfu Lu}, \bibinfo{person}{Yuan Fang},
  {and} \bibinfo{person}{Chuan Shi}.} \bibinfo{year}{2020}\natexlab{}.
\newblock \showarticletitle{Meta-learning on heterogeneous information networks
  for cold-start recommendation}. In \bibinfo{booktitle}{\emph{Proceedings of
  the 26th ACM SIGKDD International Conference on Knowledge Discovery \& Data
  Mining}}. \bibinfo{pages}{1563--1573}.
\newblock


\bibitem[Ma et~al\mbox{.}(2018)]%
        {ma2018modeling}
\bibfield{author}{\bibinfo{person}{Jiaqi Ma}, \bibinfo{person}{Zhe Zhao},
  \bibinfo{person}{Xinyang Yi}, \bibinfo{person}{Jilin Chen},
  \bibinfo{person}{Lichan Hong}, {and} \bibinfo{person}{Ed~H Chi}.}
  \bibinfo{year}{2018}\natexlab{}.
\newblock \showarticletitle{Modeling task relationships in multi-task learning
  with multi-gate mixture-of-experts}. In \bibinfo{booktitle}{\emph{Proceedings
  of the 24th ACM SIGKDD international conference on knowledge discovery \&
  data mining}}. \bibinfo{pages}{1930--1939}.
\newblock


\bibitem[Ma et~al\mbox{.}(2020)]%
        {ma2020off}
\bibfield{author}{\bibinfo{person}{Jiaqi Ma}, \bibinfo{person}{Zhe Zhao},
  \bibinfo{person}{Xinyang Yi}, \bibinfo{person}{Ji Yang},
  \bibinfo{person}{Minmin Chen}, \bibinfo{person}{Jiaxi Tang},
  \bibinfo{person}{Lichan Hong}, {and} \bibinfo{person}{Ed~H Chi}.}
  \bibinfo{year}{2020}\natexlab{}.
\newblock \showarticletitle{Off-policy learning in two-stage recommender
  systems}. In \bibinfo{booktitle}{\emph{Proceedings of The Web Conference
  2020}}. \bibinfo{pages}{463--473}.
\newblock


\bibitem[Menon et~al\mbox{.}(2021)]%
        {menon2020long}
\bibfield{author}{\bibinfo{person}{Aditya~Krishna Menon},
  \bibinfo{person}{Sadeep Jayasumana}, \bibinfo{person}{Ankit~Singh Rawat},
  \bibinfo{person}{Himanshu Jain}, \bibinfo{person}{Andreas Veit}, {and}
  \bibinfo{person}{Sanjiv Kumar}.} \bibinfo{year}{2021}\natexlab{}.
\newblock \showarticletitle{Long-tail learning via logit adjustment}. In
  \bibinfo{booktitle}{\emph{ICLR 2021}}.
\newblock


\bibitem[Niu et~al\mbox{.}(2020)]%
        {niu2020dual}
\bibfield{author}{\bibinfo{person}{Xichuan Niu}, \bibinfo{person}{Bofang Li},
  \bibinfo{person}{Chenliang Li}, \bibinfo{person}{Rong Xiao},
  \bibinfo{person}{Haochuan Sun}, \bibinfo{person}{Hongbo Deng}, {and}
  \bibinfo{person}{Zhenzhong Chen}.} \bibinfo{year}{2020}\natexlab{}.
\newblock \showarticletitle{A dual heterogeneous graph attention network to
  improve long-tail performance for shop search in e-commerce}. In
  \bibinfo{booktitle}{\emph{Proceedings of the 26th ACM SIGKDD International
  Conference on Knowledge Discovery \& Data Mining}}.
  \bibinfo{pages}{3405--3415}.
\newblock


\bibitem[Reed(2001)]%
        {reed2001pareto}
\bibfield{author}{\bibinfo{person}{William~J Reed}.}
  \bibinfo{year}{2001}\natexlab{}.
\newblock \showarticletitle{The Pareto, Zipf and other power laws}.
\newblock \bibinfo{journal}{\emph{Economics letters}} \bibinfo{volume}{74},
  \bibinfo{number}{1} (\bibinfo{year}{2001}), \bibinfo{pages}{15--19}.
\newblock


\bibitem[Ren et~al\mbox{.}(2020)]%
        {ren2020balanced}
\bibfield{author}{\bibinfo{person}{Jiawei Ren}, \bibinfo{person}{Cunjun Yu},
  \bibinfo{person}{Shunan Sheng}, \bibinfo{person}{Xiao Ma},
  \bibinfo{person}{Haiyu Zhao}, \bibinfo{person}{Shuai Yi}, {and}
  \bibinfo{person}{Hongsheng Li}.} \bibinfo{year}{2020}\natexlab{}.
\newblock \showarticletitle{Balanced meta-softmax for long-tailed visual
  recognition}.
\newblock \bibinfo{journal}{\emph{arXiv preprint arXiv:2007.10740}}
  (\bibinfo{year}{2020}).
\newblock


\bibitem[Toneva et~al\mbox{.}(2018)]%
        {toneva2018empirical}
\bibfield{author}{\bibinfo{person}{Mariya Toneva}, \bibinfo{person}{Alessandro
  Sordoni}, \bibinfo{person}{Remi Tachet~des Combes}, \bibinfo{person}{Adam
  Trischler}, \bibinfo{person}{Yoshua Bengio}, {and}
  \bibinfo{person}{Geoffrey~J Gordon}.} \bibinfo{year}{2018}\natexlab{}.
\newblock \showarticletitle{An empirical study of example forgetting during
  deep neural network learning}.
\newblock \bibinfo{journal}{\emph{arXiv preprint arXiv:1812.05159}}
  (\bibinfo{year}{2018}).
\newblock


\bibitem[Van Der~Maaten(2014)]%
        {van2014accelerating}
\bibfield{author}{\bibinfo{person}{Laurens Van Der~Maaten}.}
  \bibinfo{year}{2014}\natexlab{}.
\newblock \showarticletitle{Accelerating t-SNE using tree-based algorithms}.
\newblock \bibinfo{journal}{\emph{The Journal of Machine Learning Research}}
  \bibinfo{volume}{15}, \bibinfo{number}{1} (\bibinfo{year}{2014}),
  \bibinfo{pages}{3221--3245}.
\newblock


\bibitem[Wang et~al\mbox{.}(2020a)]%
        {wang2020improving}
\bibfield{author}{\bibinfo{person}{Kaixin Wang}, \bibinfo{person}{Bingyi Kang},
  \bibinfo{person}{Jie Shao}, {and} \bibinfo{person}{Jiashi Feng}.}
  \bibinfo{year}{2020}\natexlab{a}.
\newblock \showarticletitle{Improving generalization in reinforcement learning
  with mixture regularization}. In \bibinfo{booktitle}{\emph{NeurIPS}}.
\newblock


\bibitem[Wang et~al\mbox{.}(2019)]%
        {wang2019neural}
\bibfield{author}{\bibinfo{person}{Xiang Wang}, \bibinfo{person}{Xiangnan He},
  \bibinfo{person}{Meng Wang}, \bibinfo{person}{Fuli Feng}, {and}
  \bibinfo{person}{Tat-Seng Chua}.} \bibinfo{year}{2019}\natexlab{}.
\newblock \showarticletitle{Neural graph collaborative filtering}. In
  \bibinfo{booktitle}{\emph{SIGIR}}. \bibinfo{pages}{165--174}.
\newblock


\bibitem[Wang et~al\mbox{.}(2020b)]%
        {wang2020long}
\bibfield{author}{\bibinfo{person}{Xudong Wang}, \bibinfo{person}{Long Lian},
  \bibinfo{person}{Zhongqi Miao}, \bibinfo{person}{Ziwei Liu}, {and}
  \bibinfo{person}{Stella~X Yu}.} \bibinfo{year}{2020}\natexlab{b}.
\newblock \showarticletitle{Long-tailed recognition by routing diverse
  distribution-aware experts}.
\newblock \bibinfo{journal}{\emph{arXiv preprint arXiv:2010.01809}}
  (\bibinfo{year}{2020}).
\newblock


\bibitem[Yao et~al\mbox{.}(2021)]%
        {yao2021self}
\bibfield{author}{\bibinfo{person}{Tiansheng Yao}, \bibinfo{person}{Xinyang
  Yi}, \bibinfo{person}{Derek~Zhiyuan Cheng}, \bibinfo{person}{Felix Yu},
  \bibinfo{person}{Ting Chen}, \bibinfo{person}{Aditya Menon},
  \bibinfo{person}{Lichan Hong}, \bibinfo{person}{Ed~H Chi},
  \bibinfo{person}{Steve Tjoa}, \bibinfo{person}{Jieqi Kang}, {et~al\mbox{.}}}
  \bibinfo{year}{2021}\natexlab{}.
\newblock \showarticletitle{Self-supervised learning for large-scale item
  recommendations}. In \bibinfo{booktitle}{\emph{Proceedings of the 30th ACM
  International Conference on Information \& Knowledge Management}}.
  \bibinfo{pages}{4321--4330}.
\newblock


\bibitem[Yi et~al\mbox{.}(2019)]%
        {yi2019sampling}
\bibfield{author}{\bibinfo{person}{Xinyang Yi}, \bibinfo{person}{Ji Yang},
  \bibinfo{person}{Lichan Hong}, \bibinfo{person}{Derek~Zhiyuan Cheng},
  \bibinfo{person}{Lukasz Heldt}, \bibinfo{person}{Aditee Kumthekar},
  \bibinfo{person}{Zhe Zhao}, \bibinfo{person}{Li Wei}, {and}
  \bibinfo{person}{Ed Chi}.} \bibinfo{year}{2019}\natexlab{}.
\newblock \showarticletitle{Sampling-bias-corrected neural modeling for large
  corpus item recommendations}. In \bibinfo{booktitle}{\emph{Proceedings of the
  13th ACM Conference on Recommender Systems}}. \bibinfo{pages}{269--277}.
\newblock


\bibitem[Zhang et~al\mbox{.}(2020)]%
        {zhang2019identity}
\bibfield{author}{\bibinfo{person}{Chiyuan Zhang}, \bibinfo{person}{Samy
  Bengio}, \bibinfo{person}{Moritz Hardt}, \bibinfo{person}{Michael~C Mozer},
  {and} \bibinfo{person}{Yoram Singer}.} \bibinfo{year}{2020}\natexlab{}.
\newblock \showarticletitle{Identity crisis: Memorization and generalization
  under extreme overparameterization}. In \bibinfo{booktitle}{\emph{ICLR}}.
\newblock


\bibitem[Zhang et~al\mbox{.}(2021a)]%
        {zhang2021model}
\bibfield{author}{\bibinfo{person}{Yin Zhang}, \bibinfo{person}{Derek~Zhiyuan
  Cheng}, \bibinfo{person}{Tiansheng Yao}, \bibinfo{person}{Xinyang Yi},
  \bibinfo{person}{Lichan Hong}, {and} \bibinfo{person}{Ed~H Chi}.}
  \bibinfo{year}{2021}\natexlab{a}.
\newblock \showarticletitle{A model of two tales: Dual transfer learning
  framework for improved long-tail item recommendation}. In
  \bibinfo{booktitle}{\emph{Proceedings of the web conference 2021}}.
  \bibinfo{pages}{2220--2231}.
\newblock


\bibitem[Zhang et~al\mbox{.}(2021b)]%
        {zhang2021test}
\bibfield{author}{\bibinfo{person}{Yifan Zhang}, \bibinfo{person}{Bryan Hooi},
  \bibinfo{person}{Lanqing Hong}, {and} \bibinfo{person}{Jiashi Feng}.}
  \bibinfo{year}{2021}\natexlab{b}.
\newblock \showarticletitle{Test-agnostic long-tailed recognition by test-time
  aggregating diverse experts with self-supervision}.
\newblock \bibinfo{journal}{\emph{arXiv e-prints}} (\bibinfo{year}{2021}),
  \bibinfo{pages}{arXiv--2107}.
\newblock


\bibitem[Zheng et~al\mbox{.}(2021)]%
        {zheng2021disentangling}
\bibfield{author}{\bibinfo{person}{Yu Zheng}, \bibinfo{person}{Chen Gao},
  \bibinfo{person}{Xiang Li}, \bibinfo{person}{Xiangnan He},
  \bibinfo{person}{Yong Li}, {and} \bibinfo{person}{Depeng Jin}.}
  \bibinfo{year}{2021}\natexlab{}.
\newblock \showarticletitle{Disentangling user interest and conformity for
  recommendation with causal embedding}. In
  \bibinfo{booktitle}{\emph{Proceedings of the Web Conference 2021}}.
  \bibinfo{pages}{2980--2991}.
\newblock


\bibitem[Zhou et~al\mbox{.}(2020)]%
        {zhou2020bbn}
\bibfield{author}{\bibinfo{person}{Boyan Zhou}, \bibinfo{person}{Quan Cui},
  \bibinfo{person}{Xiu-Shen Wei}, {and} \bibinfo{person}{Zhao-Min Chen}.}
  \bibinfo{year}{2020}\natexlab{}.
\newblock \showarticletitle{Bbn: Bilateral-branch network with cumulative
  learning for long-tailed visual recognition}. In
  \bibinfo{booktitle}{\emph{Proceedings of the IEEE/CVF conference on computer
  vision and pattern recognition}}. \bibinfo{pages}{9719--9728}.
\newblock


\bibitem[Zhou et~al\mbox{.}(2011)]%
        {zhou2011functional}
\bibfield{author}{\bibinfo{person}{Ke Zhou}, \bibinfo{person}{Shuang-Hong
  Yang}, {and} \bibinfo{person}{Hongyuan Zha}.}
  \bibinfo{year}{2011}\natexlab{}.
\newblock \showarticletitle{Functional matrix factorizations for cold-start
  recommendation}. In \bibinfo{booktitle}{\emph{Proceedings of the 34th
  international ACM SIGIR conference on Research and development in Information
  Retrieval}}. \bibinfo{pages}{315--324}.
\newblock


\bibitem[Zhu et~al\mbox{.}(2021)]%
        {zhu2021understanding}
\bibfield{author}{\bibinfo{person}{Sicheng Zhu}, \bibinfo{person}{Bang An},
  {and} \bibinfo{person}{Furong Huang}.} \bibinfo{year}{2021}\natexlab{}.
\newblock \showarticletitle{Understanding the Generalization Benefit of Model
  Invariance from a Data Perspective}. In \bibinfo{booktitle}{\emph{NeurIPS}}.
\newblock


\end{thebibliography}

\end{document}